\UseRawInputEncoding
\documentclass[aps,prl,twocolumn,showpacs,superscriptaddress]{revtex4-2}

\usepackage{stackrel,amssymb}
\usepackage{amsmath,chemarrow}

\usepackage{graphicx}

\usepackage{color, soul}

\newcommand{\prmr}{{\rm{p}}}
\newcommand{\acrcy}{\mathcal{A}}

\begin{document}

\title{Error Catastrophe {Can Be} Avoided by Proofreading Innate to Template-Directed Polymerization}

\affiliation{Simons Centre for the Study of Living Machines, National Centre for Biological Sciences, Bellary Road, Bangalore 560 065, Karnataka, India}
\affiliation{School of Biological Sciences, University of Auckland, Private Bag 92019, Auckland 1142, New Zealand}
\affiliation{Research Center for Complex Systems Biology, Universal Biology Institute, University of Tokyo, Komaba 3-8-1, Meguro-ku, Tokyo 153-8902, Japan}

\author{Yoshiya J. Matsubara}
\email{yoshiyam@ncbs.res.in}
\affiliation{Simons Centre for the Study of Living Machines, National Centre for Biological Sciences, Bellary Road, Bangalore 560 065, Karnataka, India}
\author{Nobuto Takeuchi}
\email{nobuto.takeuchi@auckland.ac.nz}
\affiliation{School of Biological Sciences, University of Auckland, Private Bag 92019, Auckland 1142, New Zealand}
\affiliation{Research Center for Complex Systems Biology, Universal Biology Institute, University of Tokyo, Komaba 3-8-1, Meguro-ku, Tokyo 153-8902, Japan}
\author{Kunihiko Kaneko}
\email{kaneko@complex.c.u-tokyo.ac.jp}
\affiliation{Research Center for Complex Systems Biology, Universal Biology Institute, University of Tokyo, Komaba 3-8-1, Meguro-ku, Tokyo 153-8902, Japan}

\date{\today}

\begin{abstract}
An important issue for the origins of life is ensuring the accurate maintenance of information in replicating polymers in the face of inevitable errors. Here, we investigated how this maintenance depends on reaction kinetics by incorporating the elementary steps of polymerization into the population dynamics of polymers. We found that template-directed polymerization entails an inherent error-correction mechanism akin to kinetic proofreading, generating long polymers that are more tolerant to an error catastrophe. Because this mechanism does not require enzymes, it is likely to operate under broad prebiotic conditions.
\end{abstract}

\maketitle

Template-directed polymerization is a fundamental chemical reaction for the sustained evolution of prebiotic systems. However, as in any chemical reaction, polymerization is subject to thermodynamically inevitable errors. Eigen \cite{eigen1971selforganization} investigated the impact of such errors on the population dynamics of replicating polymers. Using the so-called quasi-species (QS) model, Eigen \cite{eigen1971selforganization} showed that if the rate of error exceeds a certain threshold (i.e., \textit{error threshold}), polymers are unable to hold information in their sequence because of the combinatorial explosion of incorrect (i.e., mutant) sequences, which compete with a correct (i.e., master) sequence, a phenomenon coined ``error catastrophe'' \cite{swetina1982self,  leuthausser1987statistical, tarazona1992error, franz1997error, saakian2006exact, wagner2010second, takeuchi2012evolutionary, domingo2016quasispecies}.

Error catastrophe poses a serious issue for the origins of life, because prebiotic systems most likely lack sophisticated error-correction mechanisms. By contrast, cells possess energy-driven mechanisms such as kinetic proofreading (KPR) \cite{hopfield1974kinetic, ninio1975kinetic, bennett1979dissipation, murugan2012speed, sartori2013kinetic, pigolotti2016protocols, pineros2020kinetic,galstyan2020proofreading} that increase the accuracy of templated replication beyond expectations from free energy differences between correct and incorrect monomer pairs \cite{eigen1971selforganization, hopfield1974kinetic, andrieux2008nonequilibrium, ouldridge2017fundamental}. However, such mechanisms require multiple evolved enzymes, making them likely to be unavailable in prebiotic systems. Therefore, it remains unknown whether  a proofreading mechanism {would be} capable of operating under prebiotic conditions. 

In this letter, we propose a prebiotic proofreading mechanism based on positive feedback between the kinetics of polymerization and the population dynamics of replicating templates. We consider a kinetic model of polymerization, in which monomers are sequentially added to a primer in a template-directed manner. Using this model, we examined whether the effect of polymerization kinetics improves the tolerance of replicating templates to replication errors. We determined the condition for replicating templates to avoid an error catastrophe, and found that the accuracy of sequence information in replicating templates increases with the length of templates, in stark contrast with the prediction of Eigen's QS model.

\begin{figure}
\centering
\includegraphics[width=8cm]{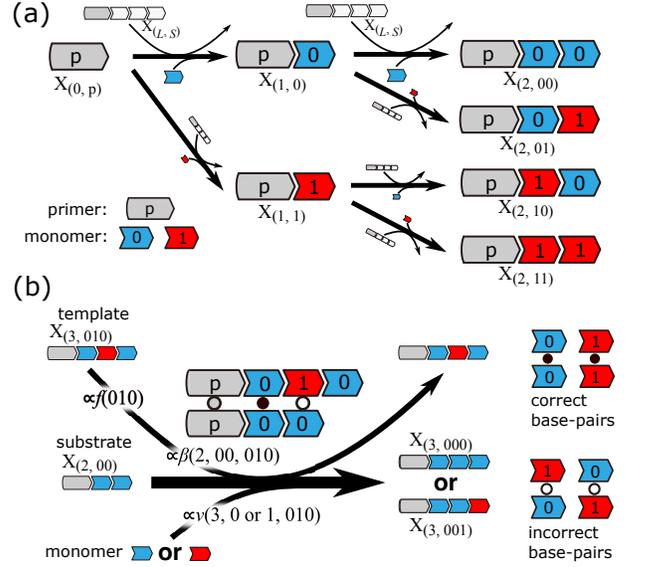}
\caption{(a) Schematic diagram of the polymerization model (up to length 2). Polymers comprise a primer (gray) and two types of monomers (blue and red). The thick arrow represents the monomer-addition reaction. In every reaction, a polymer with any sequence of length $L$ (represented as a sequence of white monomers) is used as a template. (b) Schematic of the monomer-addition reaction using a template.  In the reaction, a monomer (0 or 1) joins into a substrate polymer (e.g., $\prmr 00$), which forms pairings with a template (e.g., $\prmr 010$). The rate of the reaction is determined by the efficiency of the template $f$, the factor associated with the binding energy between the template and substrate $\beta$, and the joining rate of a monomer $\nu$. $\beta$ depends on whether the pairings between the substrate and the template are correct (filled circles) or not (open circles) (a pair of primers is depicted as a gray circle), and $\nu$ depends on the pairing between the added monomer and that at the corresponding position in the template.}
\label{fig:model}
\end{figure}

We consider a polymer (denoted by ${\rm{X}}_{l,s}$) comprising a primer (denoted by `$\prmr$') linked to a sequence of $l$ binary monomers (denoted by $s\in \{0,1 \}^l$) (e.g., `$\prmr 000$', `${\prmr} 101010$') \footnote{A primer is interpreted as a specific short polymer sequence. Primers are generally required to initiate the polymerization reaction in experimental template-directed replication, given that a monomer alone cannot bind to a template polymer.}.
The polymer is extended by the addition of a monomer (denoted by $m\in\{0,1\}$) using another polymer (denoted by $X_{L,S}$) as a template (Fig.\,\ref{fig:model}(a)):
\begin{equation*}
      {\rm{X}}_{l,s} + {\rm{X}}_{L,S} \xrightarrow{{r}(l,s,m,S)} {\rm{X}}_{l+1,sm} + {\rm{X}}_{L,S}.
\end{equation*}
For the sake of comparison with the QS model, we assume that only polymers of length $L$ can serve as templates (hereafter called templates), and that polymers cannot be longer than $L$. 
The rate of monomer addition, ${r}(l,s,m,S)$, comprises three factors (Fig.\,\ref{fig:model}(b)):
\begin{equation}
\label{eq:three-factors}
    {r}(l,s,m,S) = \beta (l,s,S) \nu (l+1,m,S) f(S).
\end{equation}

The first factor, $\beta(l,s,S)$, depends on the binding energy between ${\rm{X}}_{l,s}$ and ${\rm{X}}_{L,S}$. We assume that only a polymer bound to a template can be extended. Thus, ${r}(l,s,m,S)$ is proportional to the fraction of ${\rm{X}}_{l,s}$ bound to ${\rm{X}}_{L,S}$ denoted by $\beta(l,s,S)$, where  $\beta(l,s,S)= \exp (\hat{n}(l,s,S) \Delta)$ and where $\hat{n}(l,s,S) (\leq l)$ is the number of ‘ ‘correct'' pairs of monomers between ${\rm{X}}_{l,s}$ and ${\rm{X}}_{L,S}$ (Fig.\,\ref{fig:model}(b)). A pair is defined as correct if the monomers are of the same type \footnote{Complementarity and directionality of templates are ignored for simplicity}, and $\Delta$ ($\geq0$) is a free-energy difference between a correct and incorrect monomer pair \footnote{the Boltzmann constant multiplied by the temperature, $k_B T$, is taken to be unity.}.

The second factor in Eq.\,\ref{eq:three-factors}, $\nu(l+1,m,S)$, depends on the binding energy between the incoming monomer $m$ and the $l+1$-th monomer of the template sequence $S$. If this monomer pair is correct, $\nu(l,m,S)$ is set to $e^\Delta$; otherwise, it is set to unity \footnote{Here, we assume that the joining of a monomer to a polymer is an irreversible reaction. This situation is realized if we use an energetically activated monomer so that covalent bond formation is energetically favorable.}. We assume that both the monomer species, `0' and `1', exist at a constant concentration. Accordingly, $\mu = 1/(1+e^ \Delta)$ gives the ``error rate'' of each monomer addition in template replication.

The third factor in Eq.\,\ref{eq:three-factors}, $f(S)$, represents the efficiency of sequence $S$ as a template, which can be interpreted as the fitness of $S$. For comparison with Eigen \cite{eigen1971selforganization}, we assume a single-peak fitness landscape: $f(S) \equiv f_0$ if $S=\{0\}^L$ (denoted by $S_0$ and called the master sequence); otherwise, $f(S) \equiv f_1<f_0$ (we set $f_0= 1$ and $f_1 = 0.1$ unless otherwise noted).

We assume a chemostat condition, where a free primer (denoted by ${\rm{X}}_{0,{\prmr}}$) was supplied at rate $\phi$, and all of the chemical species were diluted at the same rate $\phi$: 
\begin{equation*}
    \varnothing  \overset{\phi}{\rightarrow} {\rm{X}}_{0,\prmr}, \qquad {\rm{X}}_{l,s} \overset{\phi}{\rightarrow} \varnothing. 
\end{equation*}
Thus, the total concentration of all primers is unity at the steady state, $ \sum_l \sum_{s \in \{0,1\}^l} x_{l,s} = 1$.
We use the boundary condition in which the concentration of ${\rm{X}}_{0,{\prmr}}$ (denoted by $x_{0,{\prmr}}$) is fixed, and the dilution rate $\phi$ is determined by $x_{0,{\prmr}}$.

In summary, the rate equation for the concentration of ${\rm{X}}_{l,s}$ (denoted by $x_{l,s}$) is
\begin{equation}
    \begin{aligned}
    \dot{x}_{l,s} =& x_{l-1,s'} \sum_{S \in \{0,1\}^L} {r}(l-1,s',m,S) x_{L,S} \\
    &- x_{l,s} \sum_{S \in \{0,1\}^L} \bigl({r}(l,s,0,S) + {r}(l,s,1,S)\bigr) x_{L,S} - \phi x_{l,s},
    \end{aligned}
        \label{eq:model}
\end{equation}
where $s'$ denotes the sequence obtained by removing the $l$-th monomer from $s$, and the dilution rate $\phi$ is determined as $\phi = \frac{x_{0,{\prmr}}}{ 1 - x_{0,{\prmr}}} (1+e^\Delta) \sum_{S \in \{0,1\}^L} f(S)  x_{L,S}$ \footnote{
By substituting the condition at the steady state $\sum_{l=1}^L \sum_{s \in \{0,1\}^l} \dot{x}_{l,s}=0$ for Eq.\,\ref{eq:model},  $\phi$ is determined as $x_{0,{\prmr}} (1+e^\Delta)\sum_{S \in \{0,1\}^L} f(S)  x_{L,S} - (1 - x_{0,{\prmr}})\phi = 0$. Note that we assume the boundary condition so that $x_{0,{\prmr}}$ is a constant.}. If $l=L$ (the maximum length), there is no second term in Eq.\,\ref{eq:model}.

In the present model, the accuracy of replicated information (denoted by $\acrcy$) is defined as the average similarity of all templates to the master sequence $S_0$ \cite{tarazona1992error}.
\begin{equation}
    \acrcy = \sum_{S \in \{0,1\}^L} \Bigl(1 - 2 \frac{h_{0,S}}{L}\Bigr) \frac{x_{L,S}}{x_L} ,
    \label{eq:acrcy}
\end{equation}
where $x_L = \sum_{S \in \{0,1\}^L} x_{L,S}$, and $h_{0,S}$ is the Hamming distance between $S_0$ and $S$. For example, $\acrcy=1$ if all the templates are the master sequence and $\acrcy=0$ if the templates are completely random sequences. 

First, we demonstrate that if the dilution rate (or, equivalently, the supply of a primer) $\phi$ is sufficiently lower than the polymerization rate (i.e., $\phi \sim 0$, and $x_{0,{\prmr}} \ll 1$), the model is reduced to the QS model. Therefore, the rate equation for $x_{L,S}$ is reduced to
\begin{equation}
    \label{eq:qsm}
    \dot{x}_{L,S} = \sum_{S'\in\{0,1\}^L} \omega_{S,S'} x_{L,S'} - \phi x_{L,S},
\end{equation}
where $\omega_{S,S'} = (1+e^\Delta) x_{0,{\prmr}} f(S') \mu^{h_{S,S'}}(1-\mu)^{L-h_{S,S'}}$, and $\mu$ is the error rate (i.e., $\mu=\frac{1}{1+e^\Delta}$) (see the Supplemental Material for the derivation \cite{NoteX}). This is because for low $\phi$, a primer is extended to the maximum length $L$ immediately after binding to a template, and the fast polymerization process does not affect the slow population dynamics of the templates.

Hence, in this low-dilution-rate limit, our model gives the same error catastrophe and threshold as given by the QS model: the accuracy of the replicated information $\acrcy$, defined as Eq.\,\ref{eq:acrcy}, decreases monotonically with the increase in $\mu$, and the decrease is accelerated as the sequence length of templates $L$ increases. Based on the correspondence of Eqs.\,\ref{eq:qsm} and the QS model, the error rate $\mu$ has to be  smaller than the error threshold in order to maintain $\acrcy \sim 1$:
\begin{equation}
    \mu \lesssim \log(W)/L,
    \label{eq:threshold}
\end{equation}
where $W \equiv \frac{f_0}{f_1}$ is the advantage of the master sequence. This is the error threshold derived by Eigen~\cite{eigen1971selforganization}.

We then consider the situation in which the dilution rate $\phi$ is comparable to the polymerization rate. In this situation, the model is not reduced to the QS model, and the polymerization kinetics significantly affect the accuracy of replicated information $\acrcy$. We computed $\acrcy$ in the steady state in Eq.\,\ref{eq:model}, varying the fixed concentration of the free primer ($x_{0, {\prmr}}$) by tuning the dilution rate $\phi$ (Fig.\,\ref{fig:accuracy4}(a))~\footnote{{The relaxation dynamics toward the steady state is discussed in the Supplemental Material}~\cite{NoteX}}. If $x_{0, {\prmr}}$ (i.e., $\phi$) is low, the accuracy $\acrcy$ approaches that of the QS model, as expected (Fig.\,\ref{fig:accuracy4}(b)). As $x_{0, {\prmr}}$ (i.e., $\phi$) increases, $\acrcy$ monotonically increases for any error rate $\mu$. In other words, slow polymerization relative to dilution improves the accuracy $\acrcy$.

\begin{figure}
\centering
\includegraphics[width=8.5cm]{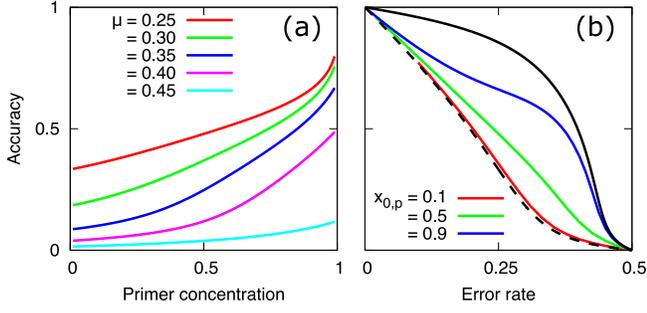}
\caption{(a) Dependence of the accuracy of the replicated information $\acrcy$ (defined as Eq.\,\ref{eq:acrcy}) upon the fixed concentration of free primers, $x_{0,{\prmr}}$ for template length $L=6$. Different error rates $\mu$ are plotted with different colors. (b) Dependence of $\acrcy$ on $\mu$ for different $x_{0,{\prmr}}$, as plotted with different colored lines. The black dashed line represents the result of the QS model, whereas the black solid line represents $\acrcy$ in the case of an infinite dilution rate, $x_{0,{\prmr}} \sim 1$, calculated by Eq.\,\ref{eq:effective}. 
}
\label{fig:accuracy4}
\end{figure}

The increased accuracy is due to the polymerization process, which works as a multistep error correction for each monomer site in the template sequence. Here, we derive the maximally achievable accuracy in the limit of an infinite dilution rate $\phi$  (i.e., $x_{0,{\prmr}} \sim 1$).  

First, we consider how the concentrations of polymers depend on the monomer at a specific site. Let the concentration of the polymers of length $l$ whose $i$-th monomer is $m\in\{0,1\}$ be expressed as $\xi_{l,m}^{(i)}x_l$, where $\xi_{l,0}^{(i)} + \xi_{l,1}^{(i)} = 1$, $x_l = \sum_{s \in \{0,1\}^l} x_{l,s}$, and $i\leq l\leq L$. Using Eq.\,\ref{eq:model}, we can show that as $\phi\to\infty$, 
\begin{equation}
    \frac{\xi_{l,0}^{(i)}}{\xi_{l,1}^{(i)}}= \frac{F_0^{(i)}}{F_1^{(i)}} \frac{\xi_{l-1,0}^{(i)}}{\xi_{l-1,1}^{(i)}},
    \label{eq:recursive}
\end{equation}
where $F_m^{(i)}$ is the relative rate at which the polymers of length $l-1$ are extended to the polymers of length $l$ whose $i$-th monomer is $m$ (see Fig.\,S1 in the Supplemental Material \cite{NoteX}). The values of $F_m^{(i)}$ are estimated as
\begin{equation}
        \begin{aligned}
        F_0^{(i)} &=& &e^{\Delta} \xi_{0} f_0 + e^{\Delta} (\xi_0^{(i)} - \xi_{0}) f_1 + \xi_1^{(i)} f_1, \\
        F_1^{(i)} &=& &\xi_{0} f_0 + (\xi_0^{(i)} - \xi_{0}) f_1 + e^{\Delta} \xi_1^{(i)} f_1,
    \end{aligned}
    \label{eq:f0f1}
\end{equation}
where $\xi_m^{(i)}=\xi_{L,m}^{(i)}$, and $\xi_{0}$ is the fraction of the master sequence, i.e., $\xi_{0} = \prod_j \xi_0^{(j)}$
(see the derivation in the Supplemental Material \cite{NoteX}).  To derive Eq.\,\ref{eq:f0f1}, we assumed that monomer additions at different positions in a sequence are independent of each other \footnote{This situation is satisfied, at least under the fitness landscape assumed in this letter}. In both lines of Eq.\,\ref{eq:f0f1}, the first term represents the rate of monomer addition using the master sequence {(i.e., the sequence with all 0s)} as a template, and the second and third terms represent rates using other templates with the $i$-th monomer 0 and 1, respectively. 

Next, polymer sequences of length $i-1$ must undergo $L+1-i$ steps of monomer-addition reactions in order to complete the synthesis of a template with length $L$. 
Using Eq.\,\ref{eq:recursive} recursively, 
the fractions $\xi_0^{(i)}$ and $\xi_1^{(i)}$ are derived by self-consistently solving
\begin{equation}
    \label{eq:effective}
    \frac{\xi_0^{(i)}}{\xi_1^{(i)}} = \frac{(F_0^{(i)})^{L+1-i}} {(F_1^{(i)})^{L+1-i}}.
\end{equation}
As shown in Fig.\,\ref{fig:accuracy4}(b), we calculated the accuracy $\acrcy$ using this estimate for $\xi_0^{(i)}$ as $\acrcy = \frac{2}{L}\sum_i \xi_0^{(i)} - 1$, which agrees well with the simulation result for $x_{0,\prmr} = 0.9$.

The effective error rate $\xi_1^{(i)}$ given by the solution of Eq.\,\ref{eq:effective} is less than the original error rate $\mu$ as described below, which suggests a proofreading effect working. By assuming the dominance of the master sequence ($\xi_0^{(i)} \sim \xi_{0} $, $ \xi_1^{(i)} \sim 0$), $\xi_1^{(i)}$ is approximated as
\begin{equation}
    \xi_1^{(i)} \sim \frac {1}{1+e^{(L+1-i) \Delta}} \sim e^{-(L+1-i) \Delta},
    \label{eq:form}
\end{equation}
because $F_0^{(i)} / F_1^{(i)}$ is approximated by $e^{\Delta}$. Eq.\,\ref{eq:form} agrees with the minimum error rate that can be achieved in the KPR model with $L+1-i$ steps, when the binding energy between the enzyme and correct/incorrect substrate differs by $\Delta$ \cite{hopfield1974kinetic, murugan2012speed}. Furthermore, even if the fraction of the master sequence $\xi_0$ is close to zero (i.e., $\Delta$ is small), the small difference between $F_0^{(i)}$ and $F_1^{(i)}$ is amplified with the powers of $L+1-i$, possibly resulting in a significant difference between $\xi_0^{(i)}$ and $\xi_1^{(i)}$, and high $\acrcy$ accordingly.

\begin{figure}
\centering
\includegraphics[width=8.2cm]{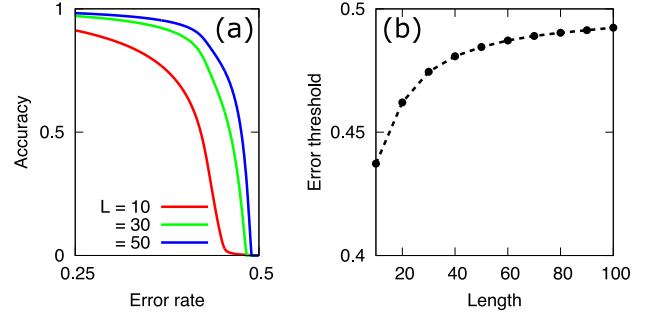}
\caption{(a) The achievable maximum accuracy of the information plotted as a function of the error rate $\mu$ ($\mu = \frac{1}{1+e^\Delta}$) for different lengths of polymer $L$ in the case with $x_{0,{\prmr}} \sim 1$. The accuracy was calculated using Eq.\,\ref{eq:effective}. (b) Dependence of the error threshold on the length of template $L$. The threshold is calculated as the error rate $\mu$, satisfying $\acrcy = 0.25$.}
\label{fig:new-threshold}
\end{figure}
 
Given the effective error rate in Eq.\,\ref{eq:effective} at each monomer site, we calculated the error threshold for the correct information to be dominant. In Fig.\,\ref{fig:new-threshold}(a), we plot the dependence of the accuracy of the information $\acrcy$ on the error rate $\mu$ with various template lengths $L$. Interestingly, the error threshold for $\mu$, at which the information is lost ($\acrcy \sim 0$), increases with the length of the template $L$ (Fig.\,\ref{fig:new-threshold}(b)). This is in sharp contrast with the QS model, where the fraction $\acrcy$ declines sharply with $L$, and the error threshold for $\mu$ approaches zero with an increase in $L$, as expressed by Eq.\,\ref{eq:threshold}.

This increase in accuracy with length $L$ is achieved, because increasing $L$ increases the number of reaction steps a monomer site in the sequence undergoes before completion of template synthesis. The effective error rate at each monomer site in the template is exponentially reduced with the number of steps, as in multistep KPR. 
Although the variety of incorrect sequences increases exponentially with $L$, as in the QS model, this is overcome by the proofreading effect (see the Supplemental Material~\cite{NoteX}).

\begin{figure}
\centering
\includegraphics[width=6cm]{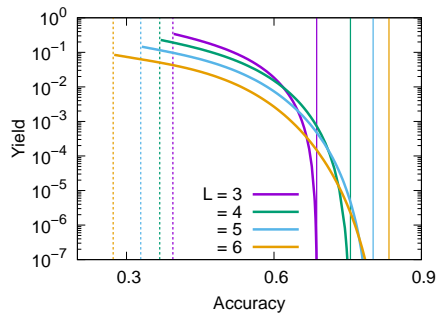}
\caption{Trade-off between the accuracy of the replicated information $\acrcy$ and the yield of the template defined as the master sequence concentration $x_{L,0}$. The solid curves represent the dependence of $x_{L,0}$ on $\acrcy$. The dilution rate $\phi$ is varied by controlling the fixed concentration of the free primer $x_{0,{\prmr}}$ in the chemostat, from $\sim 0$ to $\sim 1$. The dashed vertical lines represent the accuracy $\acrcy$ in the QS model and the solid vertical lines represent the case with $x_{0,{\prmr}} \sim 1$. We set $\Delta = 1$ (i.e., $\mu = {1}/(1+ e)$). 
}
\label{fig:trade-off}
\end{figure}

Finally, we discuss the trade-off relationship between the accuracy and yield of the templates. This trade-off is inevitable, because the accuracy of the KPR is generally achieved at the expense of synthesis efficiency~\cite{bennett1979dissipation,murugan2012speed}. In our model, we computed the yield as the actual concentration of the master sequence $x_{L,0}$. In Fig.\,\ref{fig:trade-off}, the yield is plotted against the accuracy of information $\acrcy$ by varying the dilution rate $\phi$. With an increase in $\phi$, the accuracy increases, but the yield decreases. {A similar trade-off also exists between the accuracy and energy influx (see Supplemental Material}~\cite{NoteX}).

In summary, in our template-polymerization system, the proofreading effect reduces the effective error rate as long as the dilution is not too slow. The effective error rate decreases with the template length (in sharp contrast with the QS model) in the slow dilution limit, where the error increases with the length. This proofreading effect entails a trade-off between the accuracy of replicated information and the production yield, as in the KPR scheme.

We made the following four major assumptions to make our model comparable with the QS model. {However, most of these assumptions can be relaxed for proofreading to work in our model.}

First, {we assumed immediate separation of the templates after complete full-length polymerization, as is also assumed in the QS model; thus, the separation of shorter sequences is also immediate based on thermodynamic reasoning.} {Without this assumption, the so-called ``product inhibition'' problem exists, as already pointed out for the QS model} 
~\cite{varga1997extremum, wills1998selection}. {However, the product inhibition problem could be circumvented by assuming the weak binding energy of monomer pairing while maintaining accurate replication by the proofreading effect discussed in this letter, and there are physicochemical conditions that are free from this problem}
\footnote{{We are addressing polymerization- and template-replication dynamics with product inhibition in a separate manuscript, which is in preparation.}}.  Experimentally, this condition can be realized by {the mode of driving the separation of the templates (e.g., the fast environmental, thermal or tidal, oscillation} 
 \cite{fernando2007stochastic, obermayer2011emergence, tkachenko2015spontaneous, toyabe2019cooperative}.

Second, only the longest polymers were assumed to work as templates. 
 {However, this assumption is not essential in contrast to the QS model if the shorter templates rebind to the longer templates and are consumed to produce them}~\footnote{{In the QS model, if shorter polymers of length $L'$ could work as well as long templates, the shorter templates would replicate faster and thus out-compete the longer ones. In this case, the length $L$ of a template can be replaced by a smaller $L'$, representing the so-called Spiegelman's monster problem}~\cite{spiegelman1965synthesis}.}{. Even if the shorter polymers also act as templates, our results do not significantly differ over certain parameter regions (see Supplemental Material}~\cite{NoteX}{). A similar effect has also been observed in experimental templated ligation systems}~\cite{toyabe2019cooperative}.
Because the proofreading in our model works better for a larger $L$, our results suggest that a mechanism for selecting longer polymers (e.g., \cite{kreysing2015heat, mizuuchi2019mineral}) would resolve the error-catastrophe problem because of the proofreading effect.

Third, we considered the simplest ``fitness landscape'' $f(S)$, in which only the master sequence has high fitness. {Here, ``fitness'' represents the efficiency of a template for the incorporation of monomers.} It is also possible to consider the arbitrary fitness landscape in our model (e.g., a multimodal or more rugged landscape) \cite{tarazona1992error, saakian2006exact}.  We {examined a few alternate landscapes, which supported} that the proofreading effect is relevant to avoid the error catastrophe {(see the Supplemental Material for details}~\cite{NoteX})
~\footnote{{Note that} in multimodal landscapes, the system could show multi-stability because of frequency-dependent selection among templates \cite{anderson1983suggested, matsubara2018kinetic, toyabe2019cooperative}.}.

{Fourth, we assumed that sequences always bind to templates at the same position. However, even if sequences could rebind to any site in a template, our proofreading  mechanism still holds; that is, rebinding to wrong locations does not matter because such cases are rare due to a small number of correct pairings. In most cases, a polymer extends by binding the correct site resulting in the formation of more correct pairings.}

In principle, our scheme works, even in synthetic replicating systems without complex reaction pathways such as a non-enzymatic primer-extension system \cite{rajamani2010effect, prywes2016nonenzymatic} or a template-directed ligation system \cite{toyabe2019cooperative} (see \cite{joyce2018protocells, le2019templated} for reviews). In closing, we briefly compare other schemes with the proposed model. The standard KPR currently used in biological systems requires the specific design of the reactions at each monomer addition during the replication process: a reaction pathway involving several intermediate states associated with polymerases \cite{hopfield1974kinetic} or a reverse reaction catalyzed specifically by exonucleases \cite{bennett1979dissipation}. Recently, proofreading based on a detailed polymerization mechanism coupled with cyclic {protocols} was proposed \cite{goppel2021kinetic}. By contrast, our scheme is based on general thermodynamics and the multistep nature of template replication~\footnote{{With respect to multistep polymerization, it was reported that the influence of a single mismatch can be magnified through the polymerization RecA protein in the homology search}~\cite{bar2002protein,tlusty2004high,sagi2006high}.}. The error-correction effect works at each polymerization step, which is reinforced by the positive feedback from the template population, thus enabling long templates to avoid error catastrophe. This model can therefore serve as a guide for the design of accurate template-replication systems and can further provide a plausible scenario for the inheritance of sequence information in the prebiotic world.

\begin{acknowledgments}
We thank Tetsuhiro S. Hatakeyama, Atsushi Kamimura, and Shoichi Toyabe for fruitful discussions. This research was partially supported by the Japanese Society for the Promotion of Science (JSPS) KAKENHI Grant No. 17J07169 [to Y.J.M.]; the Simons Foundation [to Y.J.M.]; a Grant-in-Aid for Scientific Research on Innovative Areas (17H06386) from the Ministry of Education, Culture, Sports, Science, and Technology of Japan [to K.K.]; and a Grant-in-Aid for Scientific Research (A)20H00123 from the JSPS [to K.K.].
\end{acknowledgments}

\bibliography{ref}

\end{document}


\title{Supplemental Material: Error Catastrophe Can Be Avoided by Proofreading Innate to Template-Directed Polymerization}

 \maketitle

\setcounter{equation}{0}
\def\theequation{S\arabic{equation}}
\setcounter{figure}{0}
\def\thefigure{S\arabic{figure}}

\subsection{Reduction of the model into the QS model in a case with slow dilution ($x_{0,{\prmr}} \sim 0$)}

We assume that polymerization is completed on a much faster time scale than that for the dilution (and the supply of primers) (i.e., $\phi \sim 0$, which is realized when $x_{0,{\prmr}} \ll 1$). Because the last term is negligible in Eq.\,2, using the steady-state condition $\dot{x}_{l,s}=0$ allows adiabatic elimination of variables $x_{l,s}$ where $l \leq L-1$ as
\begin{equation}
    x_{l,s} = x_{l-1,s'} \frac{\sum_{S \in \{0,1\}^L} f(S) \beta(l-1, s', S) \nu(l,m,S) x_{L,S}}{(1+e^\Delta)\sum_{S \in \{0,1\}^L} f(S) \beta(l, s, S) x_{L,S}}.
\end{equation}
Because $s'm$ and $s$ are identical sequences in Eq.\,2, it follows that $\beta(l-1, s', S) \nu(l,m,S) = \beta(l,s,S)$. Therefore,
\begin{equation}
    x_{l,s} = x_{l-1,s'} / (1+e^\Delta).
\end{equation}
Using this equation, we can transform the rate equation for template polymers (i.e., polymers of length $L$) as follows:
\begin{equation}
    \begin{aligned}
        \dot{x}_{L,s} &= x_{L-1,s'} \sum_{S \in \{0,1\}^L} f(S) \beta(L-1, s', S) \nu(L,m,S) x_{L,S} -\phi x_{L,s}\\
        &= \frac{x_{0,\prmr}}{(1+e^{\Delta})^{L-1}} \sum_{S \in \{0,1\}^L} f(S)  \beta(L-1, s', S) \nu(L,m,S) x_{L,S} -\phi x_{L,s}.
    \end{aligned}
\end{equation}
Because $\beta(L-1, s', S) \nu(L,m,S) = \beta(L,s,S)$, it follows that
\begin{equation}
    \begin{aligned}
        \dot{x}_{L,s} &=  \frac{x_{0,\prmr}}{(1+e^{\Delta})^{L-1}} \sum_{S \in \{0,1\}^L} f(S)  \beta(L, s', S) x_{L,S} -\phi x_{L,s}\\
        &= \frac{x_{0,\prmr}}{(1+e^{\Delta})^{L-1}} \sum_{S \in \{0,1\}^L} f(S) e^{\hat{n}(l,s,S)\Delta} x_{L,S} -\phi x_{L,s}\\
        &= x_{0,\prmr}(1+e^{\Delta}) \sum_{S \in \{0,1\}^L} f(S) \frac{e^{(L-h_{s,S})\Delta}}{(1+e^\Delta)^L} x_{L,S} -\phi x_{L,s},\\
    \end{aligned}
\end{equation}
where $h_{i,j}$ is the Hamming distance between sequences $i$ and $j$. Using $\mu=\frac{1}{1+e^\Delta}$, Eq.\,4 in the main text is obtained.

\subsection{Derivation of the upper limit of the accuracy of the information in a case with fast dilution ($x_{0,{\prmr}} \sim 1$)}

First, we assume that the frequencies of `0' and `1' at different locations along the polymers are independent of each other. Let $ \xi_{l,0}^{(i)}$ and $ \xi_{l,1}^{(i)}$ denote the relative frequencies of polymers of sequence length $l$ whose $i$-th bit is ‘0' and `1', respectively, where $ \xi_{l,0}^{(i)} + \xi_{l,1}^{(i)} = 1$. The concentration of polymers of sequence length $l$ and sequence $s$ is then expressed as follows:
\begin{equation}
    x_{l, s} = x_l \prod_{i=1}^l \xi_{l,m_i}^{(i)},
    \label{eq:bits}
\end{equation}
where $m_i$ is the $i$-th bit of sequence $s$, and $x_l$ is the sum of the concentrations of the polymers of sequence length $l$ ($x_l = \sum_{s \in \{0,1\}^l} x_{l,s}$). 

From the steady state of Eq.\,2, $x_{l,s}$ is calculated as follows:
\begin{equation}
    x_{l,s} = x_{l-1,s'} \sum_{S \in \{0,1\}^L} {r}(l-1,s',m,S) x_{L,S} /\phi,
\end{equation}
where we assumed that the first and last terms are dominant in Eq.\,2, because we assumed $x_{0,{\prmr}} \sim 1$ to allow a large $\phi$. $s'$ represents a sequence in which the end monomer of sequence $s$ is deleted. Here, substituting Eq.\,\ref{eq:bits} and summing all the concentrations of the sequences whose $i$-th bit is `0' gives
\begin{equation}
    \begin{split}
        \phi \sum_{s \in s_0^{(i)} } x_{l,s} =& \sum_{s \in s_0^{(i)} } x_{l-1} \prod_{j=1}^{l-1} \xi_{l-1,m_j}^{(j)} \sum_{S \in \{0,1\}^L} {r}(l-1,s',m_l,S) x_L \prod_{k=1}^L \xi_{L, M_k}^{(k)}, \\
        =& x_{l-1} x_L \sum_{s \in s_0^{(i)} } \xi_{l-1, 0}^{(i)}\prod_{j=1,j\neq i}^{l-1} \xi_{l-1, m_j}^{(j)}\sum_{S \in \{0,1\}^L} {r}(l-1,s',m_l,S) \prod_{k=1}^L \xi_{L, M_k}^{(k)},
    \end{split}
\end{equation}
where $\sum_{s \in s_0^{(i)}}$ denotes the summation of all of the sequences where the $i$-th monomer is `0', and $m_j$ and $M_j$ denote the $j$-th bits of the sequences $s$ and $S$, respectively. Note that if $l=i$, we should read $\xi_{l-1, 0}^{(i)}$ as $\xi_{l-1, 0}^{(i)}=1$. Here, we assume that $f(S) = f_0$ if $S$ is the master sequence, and $ = f_1$ if that is the other sequence. This gives
\begin{equation}
    \begin{split}
        \phi \sum_{s \in s_0^{(i)}} x_{l,s} = x_{l-1} x_L \sum_{s \in s_0^{(i)}} \xi_{l-1,0}^{(i)} \prod_{j=1,j\neq i}^{l-1} \xi_{l-1,m_j}^{(j)}
        \biggl(\sum_{S \in \{0,1\}^L} f_1 \nu(l,m_l, S) \beta(l-1,s',S) \prod_{k=1}^L \xi_{L,M_k}^{(k)} \\+ (f_0-f_1)\nu(l,m_l,S_0) \beta(l-1,s',S_0) \prod_{k=1}^L \xi_{L,0}^{(k)}\biggr),
    \end{split}
\end{equation}
By applying the definition of $\beta(l-1,s',S)$ and $\nu(l,m_l,S)$ for each pair of polymers, $x_{l,s} = x_l \prod_{i=1}^l \xi_{l,m_i}^{(i)}$ and a template $x_{L,S} = x_L \prod_{i=1}^L \xi_{l,M_i}^{(i)}$,
\begin{equation}
    \begin{split}
        = x_{l-1} x_L f_1  \xi_{l-1,0}^{(i)}(e^{\Delta} \xi_{L,0}^{(i)} + \xi_{L,1}^{(i)})\prod_{j=1,j\neq i}^{l} (e^{\Delta} \xi_{l-1,0}^{(j)} \xi_{L,0}^{(j)}+ \xi_{l-1,0}^{(j)} \xi_{L,1}^{(j)}+ \xi_{l-1,1}^{(j)} \xi_{L,0}^{(j)} + e^{\Delta} \xi_{l-1,1}^{(j)} \xi_{L,1}^{(j)}) \\
        + x_{l-1} x_L (f_0-f_1) e^\Delta \xi_{l-1,0}^{(i)}\prod_{k=1}^{L} \xi_{L,0}^{(k)} \prod_{j=1,j\neq i}^{l} (e^{\Delta} \xi_{l-1,0}^{(j)} + \xi_{l-1,1}^{(j)}),
    \end{split}
 \end{equation}
 where we define $\xi_{l-1,0}^{(l)} = \xi_{l-1,1}^{(l)} = 1$.
By using $\xi_{L,0}^{(i)} + \xi_{L,1}^{(i)} =1$,
 \begin{equation}
    \begin{split}
        &e^{\Delta} \xi_{l-1,0}^{(j)} + \xi_{l-1,1}^{(j)}, \\ 
         = &e^{\Delta} \xi_{l-1,0}^{(j)} (\xi_{L,0}^{(i)} + \xi_{L,1}^{(i)} ) + \xi_{l-1,1}^{(j)} (\xi_{L,0}^{(i)} + \xi_{L,1}^{(i)} ), \\ 
         = &e^{\Delta} \xi_{l-1,0}^{(j)} \xi_{L,0}^{(j)} + e^{\Delta} \xi_{l-1,0}^{(j)} \xi_{L,1}^{(j)} + \xi_{l-1,1}^{(j)} \xi_{L,0}^{(j)} + \xi_{l-1,1
        }^{(j)} \xi_{L,1}^{(j)}, \\
         = &e^{\Delta} \xi_{l-1,0}^{(j)} \xi_{L,0}^{(j)} + e^{\Delta} \xi_{l-1,0}^{(j)} \xi_{L,1}^{(j)} + \xi_{l-1,1}^{(j)} \xi_{L,0}^{(j)} + \xi_{l-1,1
        }^{(j)} \xi_{L,1}^{(j)} \\
        &+\xi_{l-1,0}^{(j)} \xi_{L,1}^{(j)} - \xi_{l-1,0}^{(j)} \xi_{L,1}^{(j)} +e^{\Delta} \xi_{l-1,1}^{(j)} \xi_{L,1}^{(j)}-e^{\Delta} \xi_{l-1,1}^{(j)} \xi_{L,1}^{(j)},\\
        = & (e^{\Delta} \xi_{l-1,0}^{(j)} \xi_{L,0}^{(j)} +  \xi_{l-1,0}^{(j)} \xi_{L,1}^{(j)}+ \xi_{l-1,1}^{(j)} \xi_{L,0}^{(j)} + e^{\Delta} \xi_{l-1,1}^{(j)} \xi_{L,1}^{(j)} ) + ( 1 - e^{\Delta} ) ( \xi_{l-1,1}^{(j)} - \xi_{l-1,0}^{(j)} ) \xi_{L,1}^{(j)}.\\
    \end{split}
 \end{equation}
In the last line, we assume that the last term is much smaller than the first term, because $( 1 - e^{\Delta} ) ( \xi_{l-1,1}^{(j)} - \xi_{l-1,0}^{(j)} ) \sim 0$ if $\Delta$ is small, and $\xi_{L,1}^{(j)} \sim 0$ if $\Delta$ is large; thus, 
\begin{equation}
    e^{\Delta} \xi_{l-1,0}^{(j)} + \xi_{l-1,1}^{(j)} \sim (e^{\Delta} \xi_{l-1,0}^{(j)} \xi_{L,0}^{(j)} +  \xi_{l-1,0}^{(j)} \xi_{L,1}^{(j)}+ \xi_{l-1,1}^{(j)} \xi_{L,0}^{(j)} + e^{\Delta} \xi_{l-1,1}^{(j)} \xi_{L,1}^{(j)} ). 
\end{equation}

\begin{figure}
\centering
\includegraphics[width=12cm]{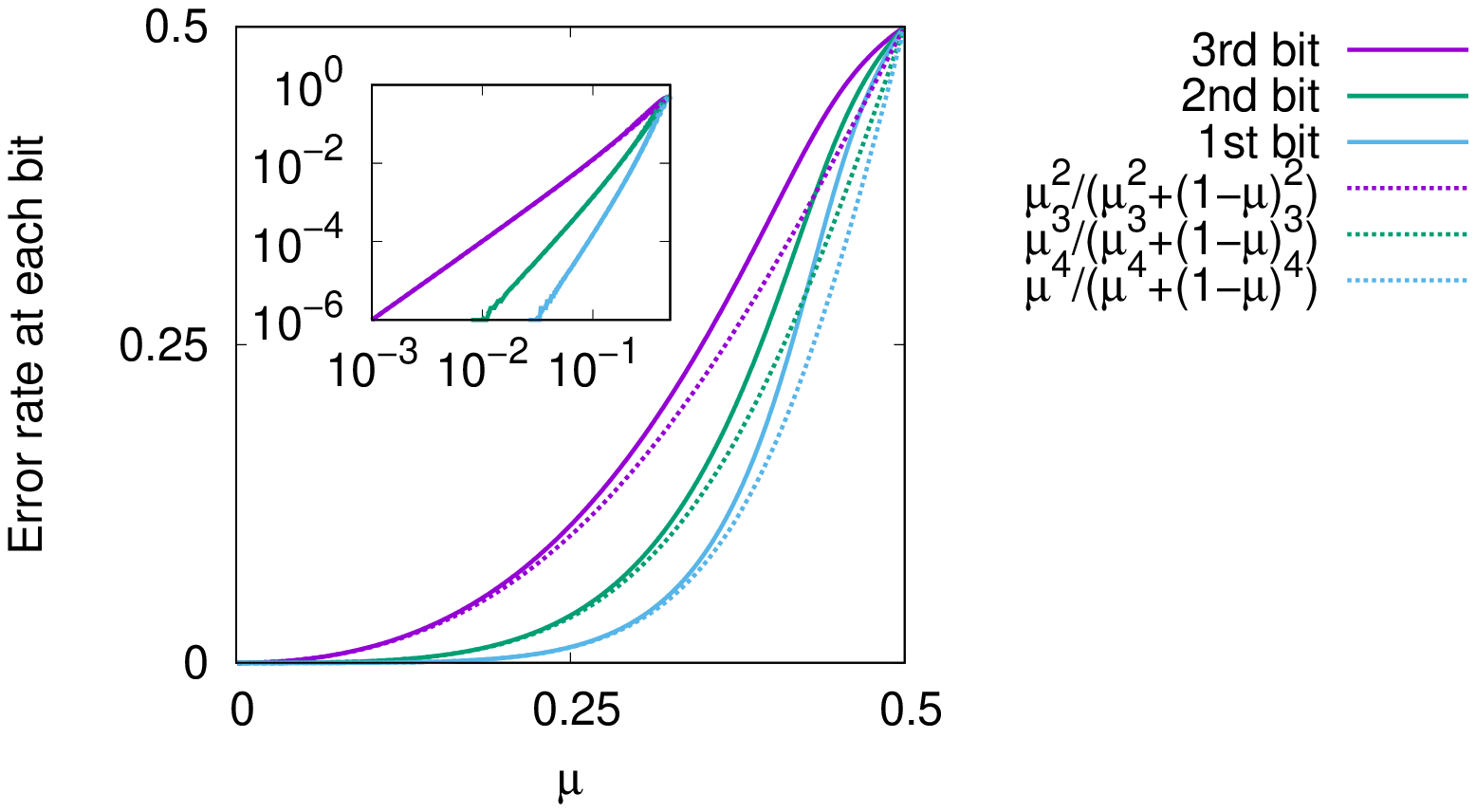}
\caption{The fraction of the error at each bit in the template sequence with length $L = 4$. The solid lines represent the fraction of the template polymer, $\xi_1^{(1)}$, $\xi_1^{(2)}$, and $\xi_1^{(3)}$. The dotted lines represent the minimum error rate that can be achieved in the KPR scheme of steps 2, 3, and 4, respectively. The inset shows the same plot with log-log axes.
}
\label{fig:error}
\end{figure}

Similarly, we obtain the expressions for the sequences whose $i$-th monomer is `1', $ \sum_{s \in s_1^{(i)}} x_{l,s}$. Thus, the relative production rate of a polymer with $i$-th monomer `0' and `1' is given by
\begin{equation}
    \begin{split}
        \phi \sum_{s \in s_0^{(i)}} x_{l,s} = \phi x_l \xi_{l,0}^{(i)} &= {A}_l^{(i)} [e^\Delta \xi_{0} f_0 + e^\Delta ( \xi_0^{(i)} -\xi_{0})f_1 + \xi_1^{(i)}f_1 ] x_{l-1} \xi_{l-1,0}^{(i)}, \\
        \phi \sum_{s \in s_1^{(i)}} x_{l,s} = \phi x_l \xi_{l,1}^{(i)} &= {A}_l^{(i)} [\xi_{0} f_0 + (\xi_0^{(i)} -\xi_{0})f_1 + e^\Delta \xi_1^{(i)}f_1 ] x_{l-1} \xi_{l-1,1}^{(i)},
    \end{split}
\end{equation}
where ${A}_l^{(i)}$ is a constant that satisfies ${A}_l^{(i)} = x_L \prod_{j=1,j\neq i}^{l} ( e^{\Delta} \xi_{l-1,0}^{(j)} + \xi_{l-1,1}^{(j)})$. Here, we define $F_0^{(i)}$ and $F_0^{(i)}$ as 
\begin{equation}
    \begin{split}
        F_0^{(i)} &= e^{\Delta} \xi_{0} f_0 + e^{\Delta} (\xi_0^{(i)} - \xi_{0}) f_1 + \xi_1^{(i)} f_1, \\
        F_1^{(i)} &=\xi_{0} f_0 + (\xi_0^{(i)} - \xi_{0}) f_1 + e^{\Delta} \xi_1^{(i)} f_1,
    \end{split}
\end{equation}
respectively, which are interpreted as the relative rates of the monomer addition to the sequence whose $i$-th monomer is `0' and `1', respectively (Fig.\,S1). Note that $F_0^{(i)}$ and $F_0^{(i)}$ do not depend on the length of the sequence. Recursive application of this process allows the fraction of the template sequence with $i$-th `0' or `1' monomer, $\xi_0^{(i)}$ or $\xi_1^{(i)}$, to be given by a self-consistent solution of
\begin{equation}
    \frac{\xi_0^{(i)}} {\xi_1^{(i)}} =  \frac{(F_0^{(i)})^{L+1-i} } { (F_1^{(i)})^{L+1-i}},
    \label{eq:selfconsistent}
\end{equation}
as explained in the main text. The numerical solution for Eq.\,\ref{eq:selfconsistent} in a case with $L=4$ is plotted in Fig.\,\ref{fig:error}.

\begin{figure}
\centering
\includegraphics[width=16cm]{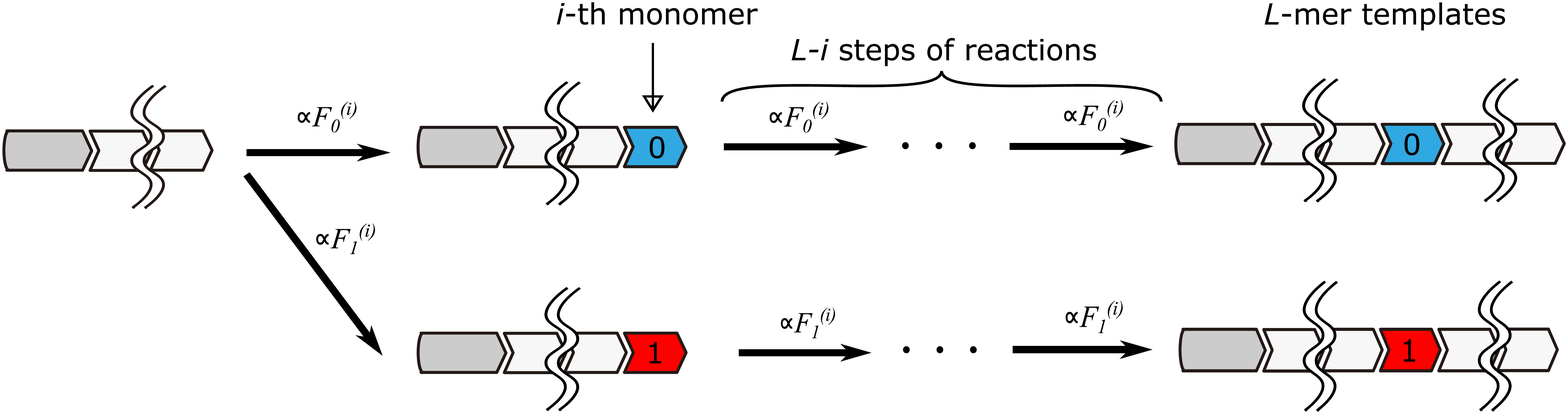}
\caption{Schematic of reaction pathways from the addition of the $i$-th monomer to the completion of the template of length $L$ whose $i$-th monomer is `0' or `1'. As in Fig.\,1(a), each arrow represents the monomer addition to a substrate polymer using a template.}
\end{figure}

\subsection{The error threshold for a replicating template with error correction}

\begin{figure}
\centering
\includegraphics[width=10cm]{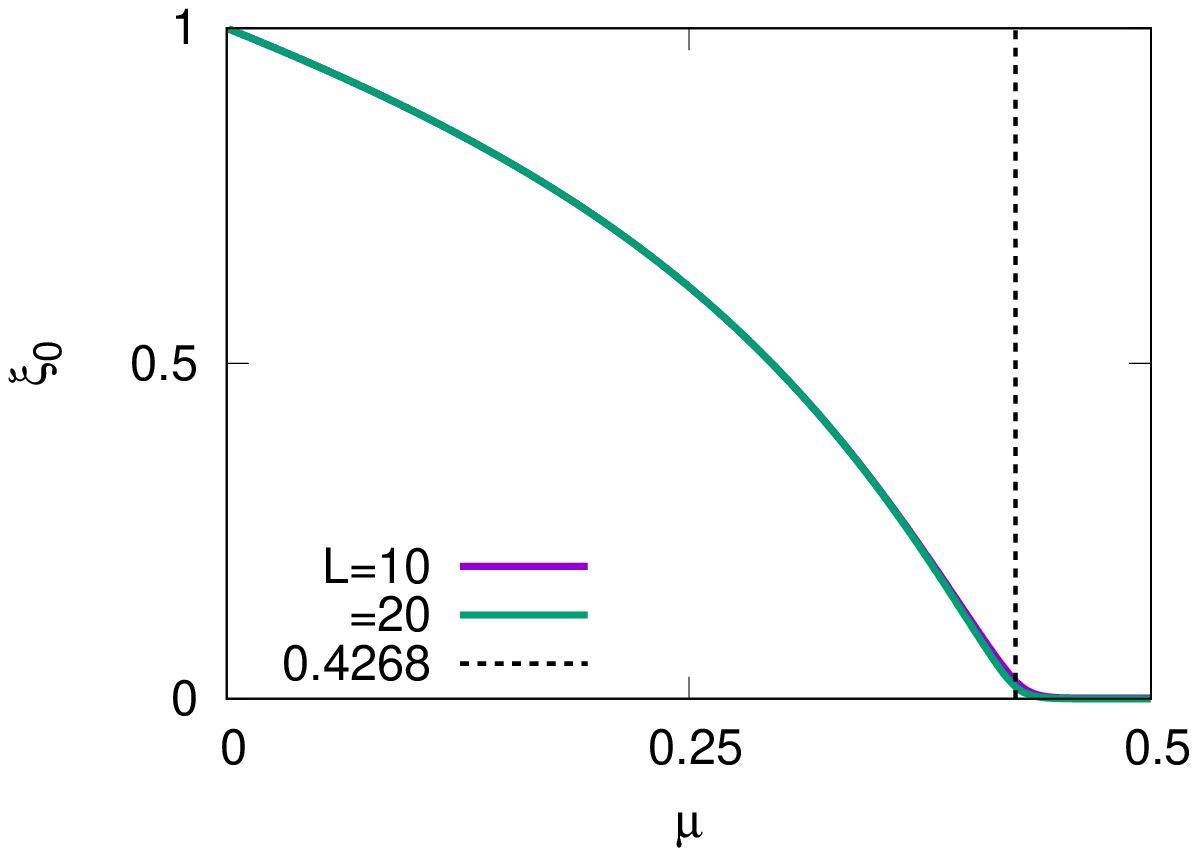}
\caption{The dependence on the error rate $\mu$ of the fraction of the master sequence among all of the templates $\xi_0$ in a case with $x_{0,{\prmr}} \sim 1$. The dashed line represents the error-rate value $\mu^*$ in the solution of Eq.\,\ref{eq:error_rate} in a case with a large $L$ limit.}
\label{fig:master_seq}
\end{figure}

The threshold value for the error catastrophe is roughly estimated in the case with a fast dilution limit (i.e., $x_{0,{\prmr}} \sim 1$).  Based on the discussion in the main text, for multistep reactions, the effective error rate at the $i$-th bit of the template during replication is modified as $\frac{\mu^{L+1-i}}{\mu^{L+1-i} + (1- \mu)^{L+1-i}}$($=\frac{1}{1+\exp(-(L+1-i)\Delta)}$). In this case, the error threshold at which the growth rate of the master sequence is overwhelmed by that of the others is estimated from the condition
 \begin{equation}
    f_0 \prod_{i=1}^L \frac{1}{1+e^{-i\Delta}}  \sim  f_1,
    \label{eq:error_rate}
 \end{equation}
where $f_0$ and $f_1$ are the fitness of the master sequence and the others, respectively. If we assume that $L$ is infinitely large, then the threshold for $\mu$ is derived numerically as $\mu^* \sim 0.4268$ (Fig.\,\ref{fig:master_seq}). It should be noted that although the fraction of the master sequence $\xi_0$ is small if $\mu > \mu^*$, the threshold for $\acrcy$ is higher than $\mu^*$, because the difference between $\xi_0^{(i)}$ and $\xi_1^{(i)}$ is magnified exponentially.
 
\subsection{Relaxation dynamics toward the steady state}

In the main text, we discussed the steady state of the templates. Here, we discuss the relaxation dynamics toward reaching such a state.

In our model, a monomer incorporation reaction does not occur without a template. Hence, long templates would not spontaneously appear if they are absent initially. Such templates are produced when including ``spontaneous ligation reactions'' from smaller monomers or polymers, as discussed previously~\cite{matsubara2016optimal, matsubara2018kinetic}. Once this occurs, even if extremely rare, the same population of polymers and templates is reached, independent of their initial concentration.

\begin{figure}[h!]
\centering
\includegraphics[width=16cm]{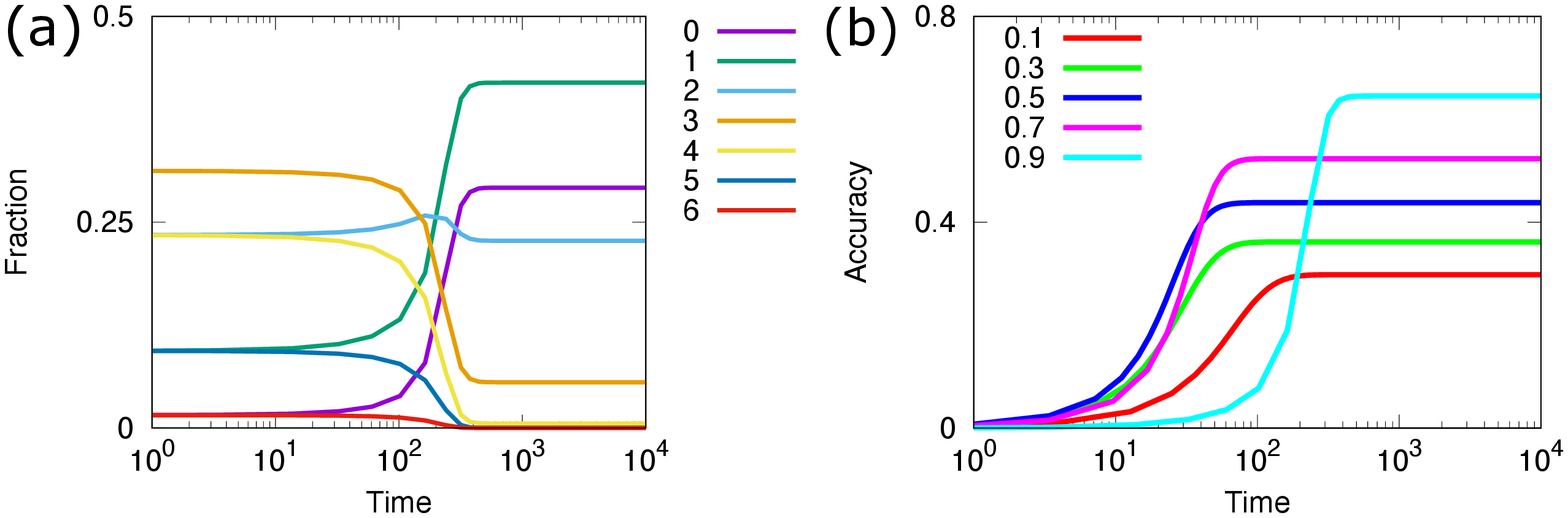}
\caption{(a) Time course of the fraction of sequences with the Hamming distance $k$ from the master sequence $S_0$, $\xi_k$. We set the concentrations of the templates to be uniform ($x_{L,S} = 10^{-4}$ for all $S$) as the initial condition. We set $L=6$, $\mu = 1/(1+e)$, and $x_{0,\prmr} = 0.9$. (b) Time course of the accuracy $\acrcy$. Each line represents the difference in $x_{0,\prmr}$. }
\label{fig:time_course}
\end{figure}

Therefore, we adopted the initial condition for the dynamics in which all template sequences exist uniformly in small amounts. We calculated the time course of the template distribution (Fig.\,\ref{fig:time_course}(a)) and the accuracy $\acrcy$ (defined in Eq.\,3) (Fig.\,\ref{fig:time_course}(b)). As expected, the dynamics eventually reached the steady state at which the master sequence is dominant, with $\acrcy \sim 1$ (if the error rate is below the threshold). Notably, the relaxation is slowed down under the fast dilution regime (i.e., large $x_\prmr$). Accordingly, this suggests a trade-off between the ``evolution speed'' and the strength of the proofreading effect.

\subsection{Energy flux to drive replication and proofreading}

For kinetic proofreading~\cite{hopfield1974kinetic} to work, energy influx is needed to drive the system toward the non-equilibrium condition; similarly, in our system, the supply of the primer and dilution corresponds to such energy influx. Then, the energy influx per template production is given as the inverse of the yield of the template, $1/x_L$ (recall that if the primer supply rate is $\phi$ and the total production rate of templates with length $L$ is $F$, then $F = x_L \phi$ at the steady state).

\begin{figure}[h!]
\centering
\includegraphics[width=16.5cm]{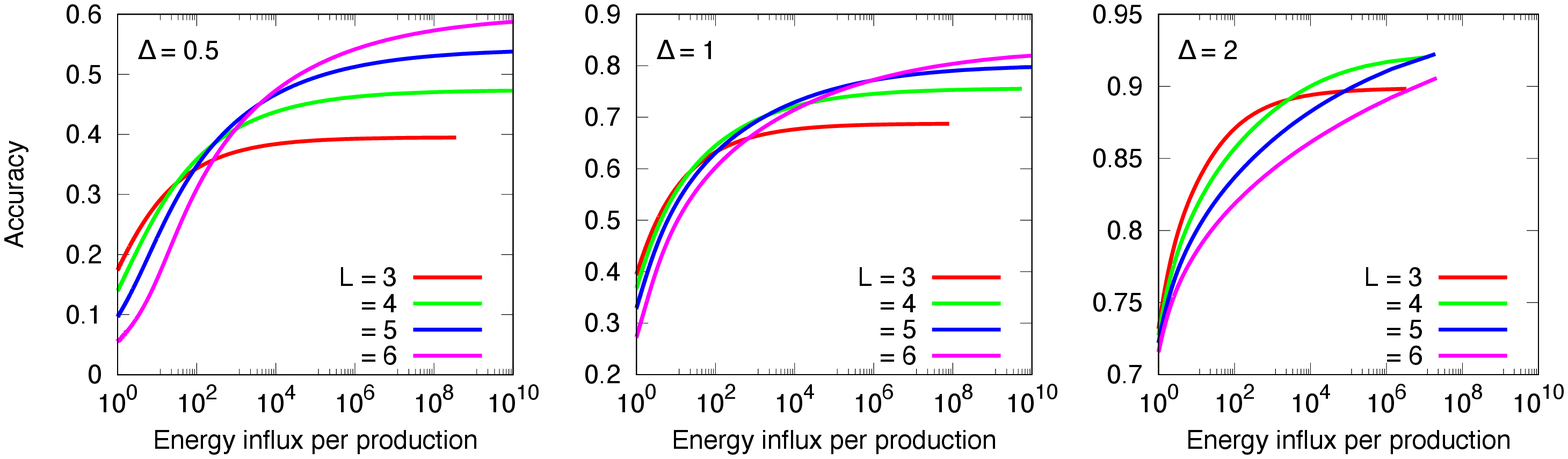}
\caption{Accuracy $\acrcy$ (defined by Eq.\,3 in the main text) versus the energy influx per template production $1/x_L$. We set $\Delta = 0.5, 1$ and $2$. }
\label{fig:energy_influx}
\end{figure}

As shown in Fig.\,\ref{fig:energy_influx}, with an increase in the energy influx (per template production), $1/x_L$, the accuracy $\acrcy$ increases until it saturates toward its maximum value. 

\subsection{Cases in which shorter polymers can also act as templates}

In the main text, we assumed that only the longest polymers with length $L$ act as templates. Here, we show that this assumption is not essential: even if shorter polymers also act as templates, the maintenance of accuracy of the information is essentially preserved.

In this revised system, the addition of a monomer using templates with arbitrary lengths is represented as
\begin{equation*}
      {\rm{X}}_{l,s} + {\rm{X}}_{L',S} \xrightarrow{} {\rm{X}}_{l+1,sm} + {\rm{X}}_{L',S},
\end{equation*}
where $l \leq L' \leq L$.
Then, we define the efficiency of sequence $S$ with length $L'$ as a template as $f(L', S)$. Here, we assume a single-peak fitness landscape: $f(L, S) \equiv f_0$ if $S=\{0\}^L$ (denoted by $S_0$), and $f(L',S) \equiv f_2$ if $ 1 \leq L' < L$; otherwise, $f(L, S) \equiv f_1<f_0$.

\begin{figure}[h!]
\centering
\includegraphics[width=15cm]{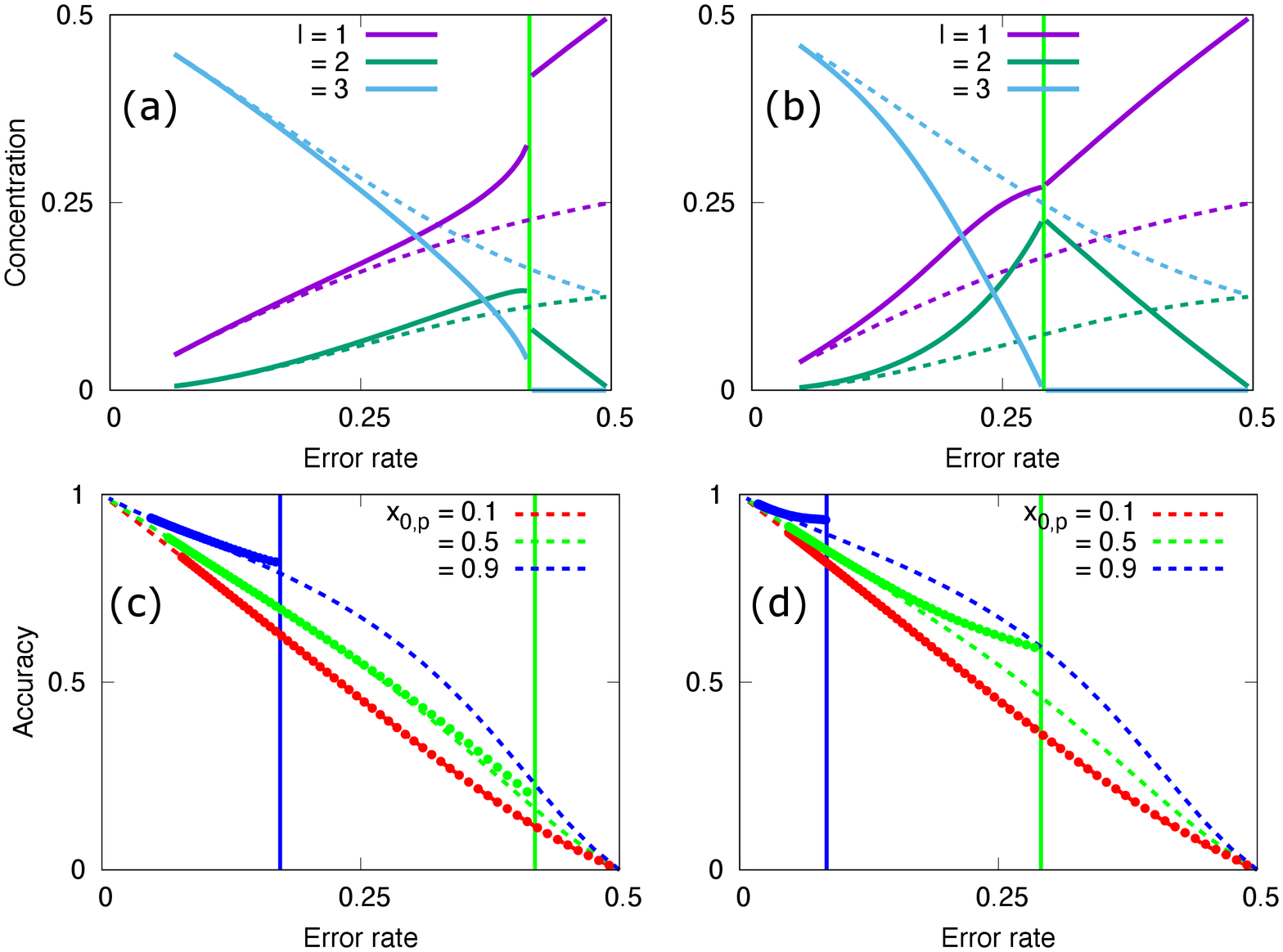}
\caption{(a, b) Total concentration of sequences with length $l$, plotted as a function of the error rate $\mu$. We set $L=3$, $x_{0,p} = 0.5$, and $f_2 = $ (a) $0.1$ or (b) $1$.  The dashed lines represent the case in which only the longest sequences with $L$ are templates (i.e., $f_2=0$), and the solid lines represent the results for the case in which shorter polymers with length $l$ ($l < L$) also work as templates. In the latter case, the longest sequences with $L$ go extinct at $\mu \sim 0.42$ in (a) and at $\sim 0.29$ in (b), as indicated by vertical lines. (c, d) The dashed lines represent the accuracy $\acrcy$ (defined as Eq.\,3) for the case in which only the longest sequences act as templates, varying the error rate $\mu$, whereas dotted lines represent the results for the case in which shorter polymers also work as templates ($f_2 = 0.1$ in (c) and $1$ in (d)). The difference in the colors represents the difference in the free primer concentration $x_{0,\prmr}$ (i.e., the dilution rate $\phi$). The vertical lines represent the error rate at which the longest sequences go extinct when $x_{0,\prmr} = 0.5$ (green) and $=0.9$ (blue). }
\label{fig:shorters}
\end{figure}

We plotted the total concentration of sequences with length $l$, $x_l = \sum_{s \in \{0,1\}^l} x_{l,s}$ by varying the error rate $\mu$ (i.e., the binding energy $\Delta$), as shown in Fig.\,\ref{fig:shorters}(a). With the increase in $\mu$ (i.e., the decrease in $\Delta$), the total concentration of the longest sequences $x_L$ decreases, and the sequences with length $L$ go extinct at a certain value of $\mu$.
Below this critical value of $\mu$, the accuracy $\acrcy$ of the information among the sequences with $L$ (defined by Eq.\,3) does not significantly differ from that found for the case in which only the sequences with $L$ act as templates (i.e., $f_2=0$), as shown in Fig.\,\ref{fig:shorters}(c). Note that even if the fitness for a shorter template $f_2$ is higher (i.e., $f_1 < f_2$), as long as $f_1 < f_0$, the results do not significantly differ, although the critical $\mu$ decreases, as shown in Fig.\,\ref{fig:shorters}(b)(d). 

\subsection{General fitness landscapes}

In the main text, we considered only a single-peaked function as the fitness landscape $f(s)$. Here, we consider more complex functions as $f(s)$, which are discussed by Tarazona~\cite{tarazona1992error}.

\subsubsection{(i) Quasi-degenerate fitness landscape}
As the first example, we investigate the following fitness function: 
\begin{equation} \label{eq:quasidegenerate}
    f(s) \equiv 
    \begin{cases}
f_0 \quad &(h_{s0} = 0) \\
f_L \quad &(h_{s0} = L) \\
f_{L-1} \quad &(h_{s0} = L-1) \\
f_1 \quad &(\text{otherwise}) \\
    \end{cases}
\end{equation}
where $h_{s0}$ is the Hamming distance from the master sequence $S_0$. 

\begin{figure}[h!]
\centering
\includegraphics[width=15cm]{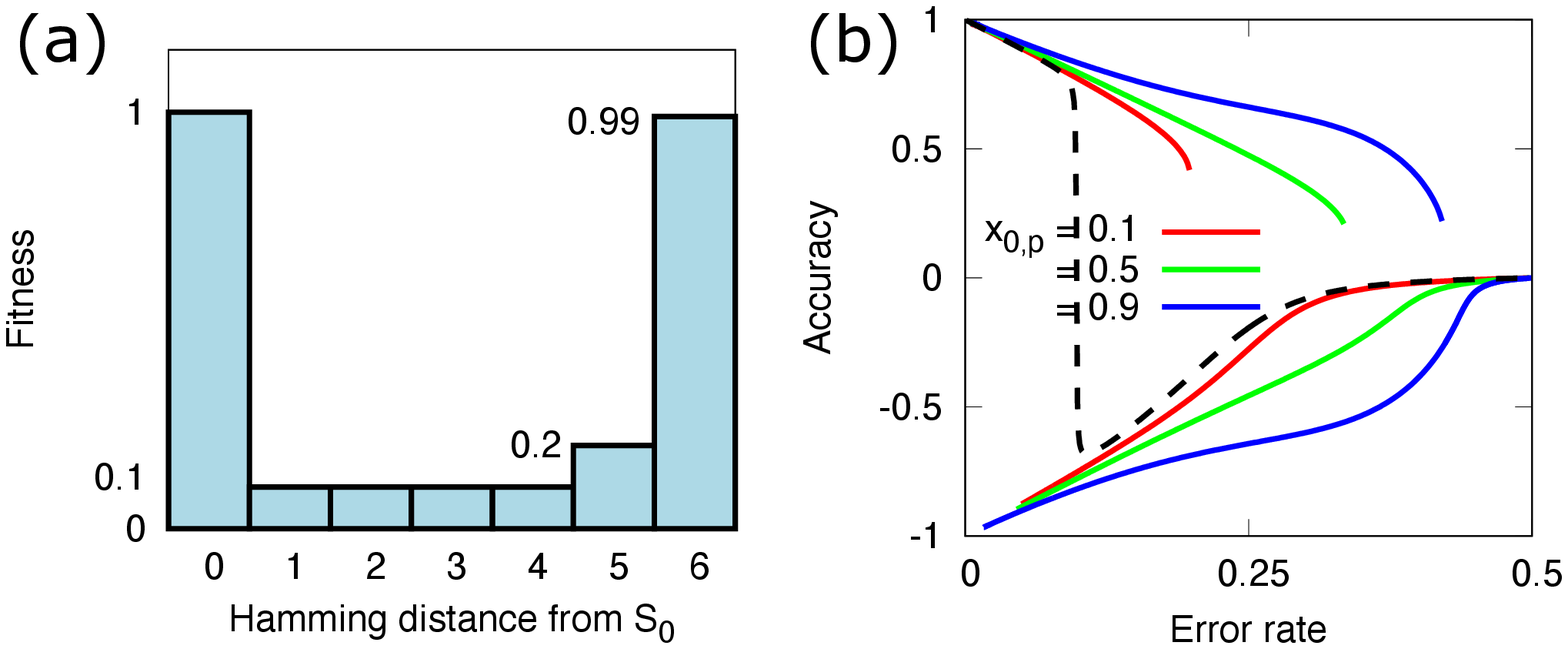}
\caption{(a) Fitness landscape for each sequence as a function of the Hamming distance from the master sequence $S_0$, defined by Eq.\,\ref{eq:quasidegenerate} when $L=6$. (b) The accuracy $\acrcy$ (defined as Eq.\,3) calculated from the steady-state concentrations of templates under the quasi-degenerate fitness landscape defined by Eq.\,\ref{eq:quasidegenerate}, plotted by varying the error rate $\mu$. The difference in the colors represents the difference in the free primer concentration $x_{0,\prmr}$ (i.e., the dilution rate $\phi$). The black dashed line represents the result derived from the quasi-species model under the same fitness landscape. We set $L=6$. }
\label{fig:quasidegenerate}
\end{figure}

If we set $f_0=1$, $f_1=0.1$, $f_L=0.99$, and $f_{L-1}=0.2$, this fitness landscape has two peaks: the sequence $S_0$ (all 0) has the highest fitness, whereas the sequence $S_L$ (all 1) has the second highest fitness with the local maximum (Fig.\,\ref{fig:quasidegenerate}(a)).

Applying this fitness landscape to the QS model (Eq.\,4 in the main text), there are three phases depending on the error rate $\mu$ (Fig.\,\ref{fig:quasidegenerate}(b))~\cite{tarazona1992error}: for low $\mu$,  the sequence $S_0$ dominates (the accuracy $\acrcy \sim 1$, as in the single-peak fitness landscape). A sudden jump occurs at a certain rate of $\mu$, and the sequence $S_L$ takes over as the dominant sequence ($\acrcy \sim -1$). The next jump (i.e., the error catastrophe) then occurs at larger $\mu$, at which the distribution of the sequences becomes uniform, and information is lost ($\acrcy \sim 0$).

Similar phases appear by varying $\mu$ when applying this fitness landscape to our polymerization model (Eq.\,2 in the main text). However, for the intermediate $\mu$ value, the steady state is not monostable but is rather bistable (Fig.\,\ref{fig:quasidegenerate}(b)); that is, both of the $S_0$-dominant ($\acrcy \sim 1$) and the $S_L$-dominant ($\acrcy \sim -1$) states are stable. Notably, the region of $\mu$ where the $S_0$-dominant state is stable expands as $x_{0,\prmr}$ increases. 
Due to the innate proofreading effect, the dominance of the (master) sequence $S_0$ is maintained even if the error rate is higher, and the error threshold is increased even under a two-peak fitness landscape.

\subsubsection{(ii) Smooth fitness landscape}

\begin{figure}[h!]
\centering
\includegraphics[width=15cm]{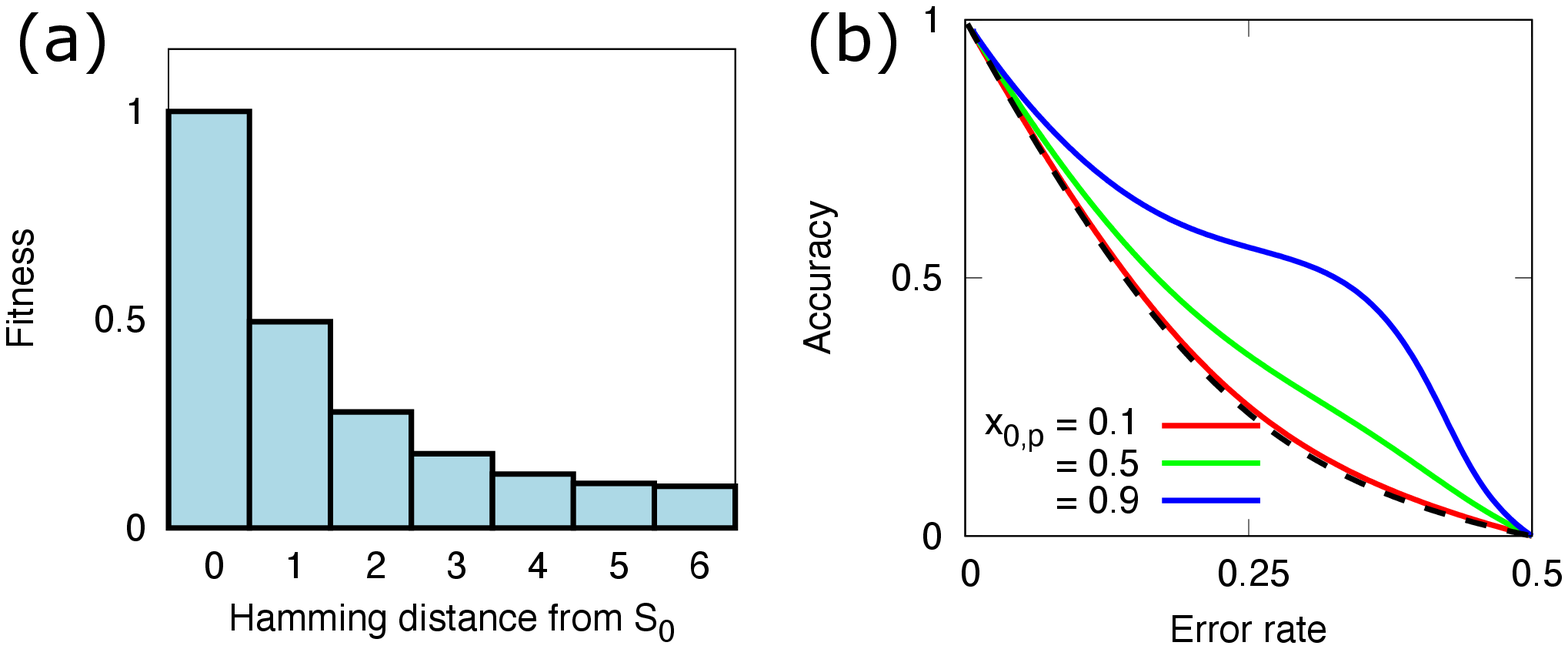}
\caption{(a) Fitness landscape for each sequence as a function of the Hamming distance from the master sequence $S_0$, defined as Eq.\,\ref{eq:smooth} when $L=6$. (b) The accuracy $\acrcy$ (defined as Eq.\,3) calculated from the steady-state concentrations of templates under the smooth fitness landscape defined as Eq.\,\ref{eq:smooth} by varying the error rate $\mu$. The difference in the colors represents the difference in the free primer concentration $x_{0,\prmr}$ (i.e., the dilution rate $\phi$). The black dashed line represents the result derived from the quasi-species model under the same fitness landscape. We set $L=6$, and $\exp (K / 2) = 10$. }
\label{fig:smooth}
\end{figure}

As the second example of an alternative fitness landscape, we consider
\begin{equation} \label{eq:smooth}
f(s) \equiv \exp\left( \frac{K}{2L^2}  \sum_{i \neq j}^L (1 - m_i) (1- m_j) \right) = \exp \left( \frac{K}{2L^2} \Bigl( (L  -   h_{0s})^2 -L  \Bigr) \right), 
\end{equation}
where $m_i$ represents the $i$-th portion of sequence $s$. This fitness landscape is inspired by the ferromagnetic Ising model, where the sequences closer to homologous sequences (i.e., all-0) have higher fitness values (see Fig.\,\ref{fig:smooth}(a) for the fitness landscape). 

Compared with the general fitness landscape assumed in the model described in the main text, this fitness landscape is not sharp, although the master sequence $S_0$ still has the highest fitness. However, as shown in Fig.\,\ref{fig:smooth}(b), the behavior of the accuracy when varying $\mu$ does not differ substantially from that under the sharp landscape (Fig.\,2(b) in the main text).

\subsubsection{(iii) Rugged fitness landscape}

As the last example, we consider the rugged landscape derived from the Hopfield model for neural networks that embeds $p$ patterns $S^k$ ($=1,\dots,p$)~\cite{hopfield1982neural}.

\begin{equation} \label{eq:rugged}
\begin{aligned}
        f(s) \equiv
          &\exp\left( \frac{1}{2L^2} \sum_k^p K_k \sum_{i \neq j}^L (1 -2 \hat m_i^k) (1-2 \hat m_j^k) (1 -2 m_i) (1-2 m_j) \right) \\
         = &\exp\left( \frac{1}{2L^2} \sum_k^p K_k \Bigl((L - 2 h_{S^k s})^2 -L  \Bigr) \right)
\end{aligned}
\end{equation}
where $\hat m_i^k$ is the $i$-th portion of the sequence $S^k$, which is randomly chosen among all sequences with length $L$. We set the master sequence $S^0 = S_0$ to have the highest fitness, and $p-1$ sequences $S^k$ also have local maximum fitness values. Recall that $m_i$ is the $i$-th portion of sequence $s$. $K_k$ is the weight of sequence $S^k$: we set $K_k = 1$ if $k=0$ and $K_k = \frac{1}{3}$ otherwise.

\begin{figure}[h!]
\centering
\includegraphics[width=8cm]{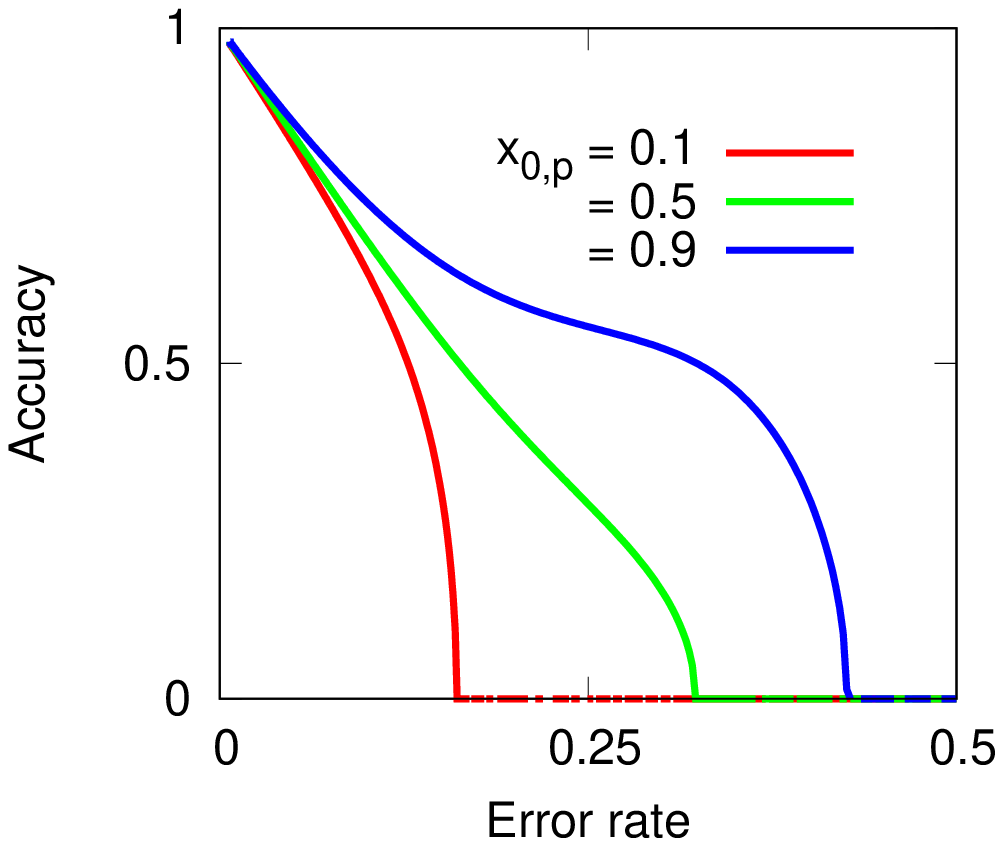}
\caption{The accuracy $\acrcy$ (defined as Eq.\,3), which maintains the state of dominance of the all-0 sequence, under the rugged fitness landscape defined as Eq.\,\ref{eq:rugged} plotted as a function of the error rate $\mu$. The colors represent the results for $x_{0,\prmr}$, the free primer concentration  (i.e., the dilution rate $\phi$). We set $L=6$. The sets of sequences with local maximum fitness $f(s)$ are given by $S^k=\{000000, 000101, 010101, 101010, 110100, 110110\}$.}
\label{fig:rugged}
\end{figure}

Under this rugged landscape, the accuracy $\acrcy$ is plotted by varying the error rate $\mu$ in Fig.\,\ref{fig:rugged}. Similar to the results obtained for the previous examples of fitness landscapes, the threshold value of $\mu$ increases with the increase in $x_{0,\prmr}$. 
This threshold increase is observed independently of the specific choice of $S^k$.

\bibliography{ref}